# Nonlinear magnetic field dependence of spin polarization in high density two-dimensional electron systems

K F Yang[1], H W Liu[1,2,5], T D Mishima[3], M B Santos[3], K Nagase[1] and Y Hirayama[1,4,5]

[1] ERATO Nuclear Spin Electronics Project, Sendai, Miyagi 980-8578, Japan

[2] State Key Lab of Superhard Materials and Institute of Atomic and Molecular Physics, Jilin University, Changchun 130012, P. R. China

[3] Homer L. Dodge Department of Physics and Astronomy, University of Oklahoma, 440 West Brooks, Norman, OK 73019-2061, USA

[4] Department of Physics, Tohoku University, Sendai, Miyagi 980-8578, Japan

[5] Author to whom any correspondence should be addressed.

E-mail: liuhw@ncspin.jst.go.jp or hirayama@m.tohoku.ac.jp

**Abstract.** The spin polarization ($P$) of high-density InSb two-dimensional electron systems (2DESs) has been measured using both parallel and tilted magnetic fields. $P$ is found to exhibit a superlinear increase with the total field $B$. This $P$-$B$ nonlinearity results in a difference in spin susceptibility between its real value $\chi_s$ and $\chi_{gm} \propto m^* g^*$ ($m^*$ and $g^*$ are the effective mass and $g$ factor, respectively) as routinely used in experiments. We demonstrate that such a $P$-$B$ nonlinearity originates from the linearly $P$-dependent $g^*$ due to the exchange coupling of electrons rather than from the electron correlation as predicted for the low-density 2DES.

**PACS:** 73.40.-c, 71.27.+a, 73.43.Qt

## 1. Introduction

The change of spin polarization ($P$) in response to an applied magnetic field ($B$) defines the spin susceptibility, a fundamental parameter in condensed matter physics. In two-dimension electron systems (2DESs) the paramagnetic spin susceptibility $\chi_0$ at $B$ = 0 plays crucial roles for the study of the 2DES ground state [1]-[9]. $\chi_0$ is usually accessible by polarizing the 2DES in a parallel or tilted $B$ [10]. In most cases, $\chi_{gm}=n_{2D}P/B \propto m^* g^*$ ($n_{2D}$ is the electron density, $m^*$ is the effective mass in units of the free electron mass $m_e$, and $g^*$ is the effective $g$ factor) obtained at finite $B$ (or $P$) is regarded as $\chi_0$. However, this attainment of $\chi_0$ has been pointed out to be valid only if $P$ is linear in $B$ [11]. More recently, theoretical calculations have predicted that $P$ is nonlinear in $B$ in 2DESs with low density (or large Wigner-Seitz radius $r_s \propto n_{2D}^{-1/2}$) due to the electron-correlation contribution [12,13]. This $P$-$B$ nonlinearity further results in a nonlinearly $P$-dependent spin susceptibility $\chi_s = n_{2D}dP/dB$, which is not equal to $\chi_{gm}$ at finite $P$. The $P$-$\chi_s$ dependence is required to investigate the partially or fully spin-polarized 2DES ground state [4,14].

Direct experimental evidence of the $P$-$B$ nonlinearity has so far been reported in low-density GaAs 2DESs with a relatively large $r_s \sim 5.6$ at $P < 0.5$ [11], where the correlation energy is believed to determine the $P$-$B$ nonlinearity [11,12]. We here, however, present a similar nonlinear $P$-$B$ dependence via a high-density InSb 2DES with a *small* $r_s \sim 0.2$. The large $g^*$ (over 39 in magnitude) of the InSb 2DES makes the $P$ = 1 state achievable at easily accessible fields and the magnetization curve ($P$ vs. its corresponding field $B_p$) observable over a wide range of $P$ from 0.07 to 1. $P$ is found to be superlinear in $B_p$, which is analogous to the findings in Ref. 11. This $P$-$B$ nonlinearity is fit well by a simple empirical equation. Note that $\chi_{gm}$ calculated by this equation can also be used to fit the non-monotonic $n_{2D}$-$\chi_{gm}$ data in Ref. 11. However, the $P$-$B$ nonlinearity in the high-density specimen does not arise from the electron correlation because of the small $r_s$. Further experiments demonstrate that this $P$-$B$ nonlinearity is attributed to a linear $P$ dependence of $g^*$.

## 2. Samples and methods

Unless otherwise noted, the employed InSb 2DES in a Hall bar (80 μm ×30 μm) is confined to a 30 nm wide InSb quantum well (QW) with $\delta$-doped $Al_{0.09}In_{0.91}Sb$ barriers on each side of the well [15]. The parallel and tilted-field measurements were performed at 200 mK in a dilution refrigerator with an *in situ* rotator. The tilt angle $\theta$ between $B$ and the sample normal (inset, figure 1(a)) was determined from the Hall resistance with an accuracy of at least 0.1°. A low-frequency AC lock-in technique (13.3 Hz, 3 nA) was used to measure the longitudinal resistance $R_{xx}$. Magnetotransport measurements gave $n_{2D} = 1.83 \times 10^{15}$ m$^{-2}$ and an electron mobility of $\mu = 14.3$ m$^2$/Vs in this sample.

## 3. Experimental results and discussion

As shown in the inset of figure 1(b), the energy space between Landau levels (LLs) with spin-up (↑) and spin-down (↓) electrons is enhanced with increasing the tilted angle. Because the cyclotron energy $E_C = \hbar e B_{perp} / m^*$ (where $\hbar$ is the reduced Planck's constant $h$, $\hbar = h/2\pi$) depends on the perpendicular component $B_{perp}$ of $B$, and the Zeeman splitting $E_Z = g^* \mu_B B = g^* \mu_B B_{perp} / \cos\theta$ ($\mu_B$ is the Bohr magneton) is dominated by $B$, the ratio $r = E_Z/E_C$ increases with increasing $\theta$ for a fixed $B_{perp}$ (or $E_C$). LLs with opposite spins intersect at integer $r$, satisfying the coincidence condition $2r\cos\theta_c = m^* g^*$ ($\theta_c$ is the coincidence angle). Note that the integer $r$ also indicates the number of fully polarized LLs. This coincidence method enables calculation of the product of $m^* g^*$ provided $r$ and $\theta_c$ are known. Moreover, $P = \Delta n/n_{2D} = r/\nu$ and the corresponding field of $B_p$ can also be determined at the LL intersection, where $\Delta n$ is the density difference between spin-up and spin-down electrons and $\nu = h n_{2D} / e B_{perp}$ is the filling factor. As a result, the LL coincidence allows us to plot the $P$-$B_p$ curve, from which $\chi_s = n_{2D} dP/dB_p$ is obtained. In addition, $\chi_{gm}$ is also given by

$$\chi_{gm} = n_{2D} \frac{P}{B_p} = \frac{e}{2h} m^* g^* \quad . \tag{1}$$

Figure 1(a) shows $R_{xx}$ versus $B_{perp}$ at several tilt angles. The trace of the peaks of $R_{xx}$ just corresponds to the LL evolution. We plot the peak positions of $R_{xx}$ as a function of $B_{perp}$ (or $\nu$) and $B$ (= $B_{perp}/\cos\theta$) in a wide range of $\theta$, as shown in figure 1(b). Two types of LL intersections are prominent.

The first is characterized by a coalescence of $R_{xx}$ minima at small $B_{perp}$ (or large $\nu$), and the other by a resistance spike [16] within persistent resistance minima (thick solid line). Symbols along the dash-dot line denote the minima and maxima of $R_{xx}$ at $\theta_c$ and integer $\nu$. We show examples of $R_{xx}$ vs. $B$ in figure 2(a). Apparently, $R_{xx}$ has a minimum (maximum) provided that both $r$ and $\nu$ have the same (different) parity (even or odd). The $R_{xx}$ minima and maxima provide precision measurements of $B_p$ at a certain $P$ and thus $m^*g^*$ based on (1). The LL intersection in this InSb 2DES is distinguishable at $r$ up to 7 and $v$ up to 14 (figure 1(b)), enabling systematic investigations of the $P$-$B_p$ dependence over a wide range of $P$ (= $r/v$) from 0.07 to 1.

Besides the data in the tilted-field region, figure 1(b) also shows data in the so-called parallel field region between $\theta = 85.3°$ and $\theta = 89.7°$. It is clear that the peak of $R_{xx}$ in this region is almost independent of $B_{perp}$. We identify the peak with the intersection between the Fermi level and the nearly degenerate upper Zeeman sublevels (thin solid line). Examples of these peaks ($\Gamma$) in the plot of $R_{xx}$ vs. $B$ is shown in figure 2(b). $R_{\mathrm{xx}}$ at low fields (left side of the peak) follows an $\sim \mathrm{e}^{B^2}$ dependence but an $\sim \mathrm{e}^{B}$ dependence at high fields (right of the dashed line) due to the spin subband depopulation [17,18]. The discontinuity in $R_{\mathrm{xx}}$ thereby leads to the appearance of the peak. Strictly speaking, the peak should be called a kink. Because Landau quantization in the InSb 2DES occurs even at small $B_{perp}$ due to the large $E_c$, depopulation of the quantized subband at $B > B_\Gamma$ (inset, figure 2(b)) is expected to suppress the intrasubband scattering of electrons and thus to reduce $R_{xx}$ until the spin subband depopulation at $B > B_c$ starts. The electron spins are fully polarized at $B > B_c$ (~ 10.8 T), as denoted by a dashed line in figure 2(b). As shown in figure 1(b), $B_{\mathrm{c}}$ determined in this way (solid square) is consistent with the critical field for the $P = 1$ state in the tilted-field region (symbols in circle). As the 2DES is tilted slightly away from $\theta = 90°$, $B_{perp}$ increases and further weakens the intrasubband scattering, resulting in a more pronounced kink.

Figure 3(a) shows the magnetization curve ($P$ vs. $B_p$, open symbols) attained from the data in figure 1(b). It is clear that $P$ is superlinear in $B_p$. Because $P$ at integer $r$ with degenerate LLs is determined by the ratio of $r/v$, its value undergoes a smooth change with $v$. This is different from the case that $P$ at $r < 1$ is always zero provided $v$ is an even integer. Therefore, $P$ at integer $r$ increases smoothly with $B_p$ ($\sim 1/v$). In other words, $B_p$ is equivalent to a parallel field. The data in figure 3(a) also give the $P$-$\chi_{gm}$ plot, as shown by the open symbols in figure 3(b). A linear fit (solid line) is calculated by

$$\chi_{gm} = \chi_0 + \Delta\chi P \quad , \tag{2}$$

where the slope $\Delta\chi = \Delta\chi_{gm}/\Delta P$ represents the strength of the spin-exchange coupling. From (1) and (2) we have

$$P = \frac{\chi_0 B_p}{n_{2D} - \Delta\chi B_p} \cdot \tag{3}$$

The thick solid line in figure 3(a) calculated from (3) is found to agree with the data. That is, this empirical equation describes well the nonlinear magnetization curve. The spin susceptibility then reads

$$\chi_{gm} = n_{2D}\frac{P}{B_p} = \chi_0 \frac{n_{2D}}{n_{2D} - \Delta\chi B_p} \quad , \tag{4}$$

$$\chi_s = n_{2D}\frac{dP}{dB_p} = \chi_0 \left( \frac{n_{2D}}{n_{2D} - \Delta\chi B_p} \right)^2 \quad . \tag{5}$$

Accordingly, we obtain

$$\frac{\chi_s}{\chi_0} = \left( \frac{\chi_{gm}}{\chi_0} \right)^2 = \left( 1 + \frac{\Delta\chi P}{\chi_0} \right)^2 \quad . \tag{6}$$

The simple law of (6) correlates $\chi_{gm}$ to the real value of the spin susceptibility $\chi_s$. This equation also indicates a nonlinear $P$-$\chi_s$ dependence, as shown by the solid line in figure 3(b). We see that $\chi_s$ is superliner in $P$, while $\chi_{gm}$ is linear in $P$. Similar results (solid symbols and dashed lines, figure 3) were also obtained from a 20 nm wide InSb QW ($n_{2D} = 3.1 \times 10^{15}$ m$^{-2}$ and $\mu = 7.9$ m$^2$/Vs) with a $\delta$-doped $Al_{0.2}In_{0.8}Sb$ barrier on one side of the well. A slight difference between the results of the two samples probably originates from different Rashba spin-orbit (SO) coupling strengths [20]. We should note that the $P$-$B_p$ nonlinearity in this work is quite similar to the observation in the low-density GaAs 2DES [11]. Furthermore, the non-monotonic $n_{2D}$-$\chi_{gm}$ curve in Ref. 11 can be fit well using our equation (4).

Because the $P$-$B_p$ nonlinearity in the GaAs 2DES occurs at relatively large $r_s$ (~ 5.6), the electron correlation is expected to contribute to this nonlinear behavior [4,11,12]. However, a similar $P$-$B_p$ nonlinearity in our InSb 2DES with a small $r_s$ (~ 0.2) allows one to question this interpretation.

From (3) we see that the $P$-$B_p$ nonlinearity is dominated by $\Delta\chi/\chi_0$. $\chi_0$ is generally accepted to increase with decreasing $n_{2D}$ (see also $\chi$ at P = 0 of the two samples in figure 3(b)), while $\Delta\chi$ exhibits diversity in various 2DESs. For examples, $\Delta\chi$ is zero in AlAs [21] and Si [22] 2DESs, suggesting a field-independent $P$ ($\chi_0 = \chi_{gm} = \chi_s$). $\Delta\chi$ is non-zero in GaAs [11], InSb [19], and CdMnTe [23] 2DESs, leading to a field-dependent $P$ ($\chi_0 \neq \chi_{gm} \neq \chi_s$ at finite $P$). As seen in figure 3(b), $\Delta\chi$ is determined by the slope of the $P$-$\chi_{gm}$ plot and accordingly determined by the $P$-dependent $m^*g^*$. Further experiment shows that the $P$-dependent $\chi_{gm}$ in our InSb 2DES arises from the $P$ dependence of $g^*$. An independent measurement of $m^*$ were performed by analyzing the temperature ($T$)-dependent amplitude ($A$) of low-field SdH oscillations (the so-called Dingle plot, figure 4(a)) [24]. The mass values $m_1$ and $m_2$ (in units of $m_e$), deduced from the Dingle plots of the first two clearly resolved SdH oscillations, give $m^* = (m_1+m_2)/2$. The attained $m^*$ (~ 0.016) is slightly larger than the mass in bulk InSb (~ 0.014) and is almost independent of $\theta$ (figure 4(b)). This result suggests a $P$-independent $m^*$. Accordingly, the $P$-enhanced $\chi_{gm}$ is attributed to the increase of $g^*$. From (2) we have

$$g^* = g^*(B=0) + \Delta g P, \qquad (7)$$

where $\Delta g$ is given by $\Delta\chi/m^*$. A significant change of $g^*$ ($\Delta g$ = 49) from 39 at $P$ = 0 to 88 at $P$ = 1 attained in our InSb 2DES is similar to the result of Ref. 19, in which the effect of Rashba SO coupling, electron-nuclear interaction, and disorder on the enhancement of $g^*$ has been addressed. In our work the $P$-$\chi_{gm}$ plots (figure 3(b)) of the two samples with different interface symmetry (corresponding to different Rashba SO coupling strengths) are similar, indicating a negligible role of the Rashba SO coupling in the greatly enhanced $g^*$. Our recent experiment has also suggested that the electron-nuclear spin coupling has no influence on enhancement of $g^*$ (or electron-spin splitting) because no nuclear polarization was detected at any LL intersection with a small current [25]. In addition, Tutuc et al. [26] have demonstrated that a large parallel field induces the increase of $m^*$ in GaAs 2DESs due to finite layer thickness [27], thereby decreasing $a_B$ and increasing $r_s$ and $g^*$. However, the tilt-angle-independent $m^*$ in our work suggests a parallel-field-independent $m^*$, excluding the role of the well thickness on $g^*$. This conclusion was also confirmed by the fact that the two samples with varied well thickness show similar results (figure 3).

From (2) and (7) we see that the enhancement of $g^*$ is dominated by the exchange energy $E_{ex}$ (proportional to the Coulomb energy $E_{coul}$) of electrons at the LL intersection. Therefore, the spin-Zeeman energy is $E_z = g^* \mu_B B_p$ with $g^* = g^*(B_p = 0) + E_{coul} P$. As mentioned above, $B_p$ in our work is equivalent to a parallel field. In parallel fields the characteristic length controlling $E_{coul}$ is related to the Fermi wave-vector $k_F$ ($E_{coul} \propto k_F$) rather than to the magnetic length $l_B$ ($E_{coul} \propto 1/l_B$) [28]. The $P$-independent $E_{coul}$ therefore gives a linearly $P$-dependent $g^*$. The ratio of the Landau width to the Zeeman splitting is relatively small in the InSb 2DES [19,29], as is expected to greatly enhance $g^*$[30,31]. At $P = 1$ the levels of spin-up and spin-down electrons separate completely, the screening effect becomes weak and thus results in a extremely large $g^*$ [31].

**Conclusions**

We have demonstrated a superlinear field dependence of spin polarization in a high-density InSb 2DES. The *P-B* nonlinearity originates from a linearly $P$-enhanced $g^*$ due to the electron exchange coupling. As a result of this *P-B* nonlinearity the real spin susceptibility $\chi_s$ is significantly larger than $\chi_{gm}$ as often used in experiments at finite $P$. Our results shed light on fundamental issue of spin susceptibility in the study of the 2DES ground state.

**Acknowledgements**

We acknowledge helpful discussions with X D Hu in University at Buffalo, R S Zak in University of Basel, and G Bauer in Delft University of Technology. H W Liu thanks the Program for New Century Excellent Talents of the University in China.

**References**

[1] Ando T, Fowler A B and Stern F 1982 *Rev. Mod. Phys.* **54** 437; Giuliani G F and Vignale G 2005 *Quantum Theory of the Electron Liquid* (Cambridge University Press, Cambridge, England)

[2] Wigner E 1934 *Phys. Rev.* **46** 1002

[3] Tanatar B and Ceperley D M 1989 *Phys. Rev.* B **39** 5005

[4] Attaccalite C, Moroni S, Gori-Giorgi P and Bachelet G B 2002 *Phys. Rev. Lett.* **88** 256601

[5] Jamei R, Kivelson S A and Spivak B 2005 *Phys. Rev. Lett.* **94** 056805; Falakshahi H and Waintal X 2005 *Phys. Rev. Lett.* **94** 046801

[6] Abrahams E, Kravchenko S V and Sarachik M P 2001 *Rev. Mod. Phys.* **73** 251; Kravchenko S V and Sarachik M P 2004 *Rep. Prog. Phys.* **67** 1

[7] Drummond N D and Needs R J 2009 *Phys. Rev. Lett.* **102** 126402

[8] Yarlagadda S and Giuliani G F 1989 *Phys. Rev.* B **40** 5432

[9] Pudalov V M, Gershenson M E, Kojima H, Butch N, Dizhur E M, Brunthaler G, Prinz A and Bauer G 2002 *Phys. Rev. Lett.* **88** 196404; Shashikin A A, Anissimova S, Sakr M R, Kravchenko S V, Dolgopolov V T and Klapwijk T M 2006 *Phys. Rev. Lett.* **96** 036403

[10] Okamoto T, Hosoya K, Kawaji S and Yagi A 1999 *Phys. Rev. Lett.* **82** 3875

[11] Zhu J, Stormer H L, Pfeiffer L N, Baldwin K W and West K W 2003 *Phys. Rev. Lett.* **90** 056805

[12] Zhang Y and Das Sarma S 2006 *Phys. Rev. Lett.* **96** 196602

[13] Subasi A L and Tanatar B 2008 *Phys. Rev.* B **78** 155304

[14] Zhang Y and Das Sarma S 2005 *Phys. Rev.* B **72** 075308

[15] Ball M A, Keay J C, Chung S J, Santos M B and Johnson M B 2002 *Appl. Phys. Lett.* **80** 2138; Chokomakoua J C, Goel N, Chung S J, Santos M B, Hicks J L, Johnson M B and Murphy S Q 2004 *Phys. Rev.* B **69** 235315

[16] Jungwirth T and MacDonald A H 2001 *Phys. Rev. Lett.* **87** 216801

[17] Yoon J, Li C C, Shahar D, Tsui D C and Shayegan M 2000 *Phys. Rev. Lett.* **84** 4421

[18] Tutuc E, Melinte S and Shayegan M 2002 *Phys. Rev. Lett.* **88** 036805.

[19] Nedniyom B, Nicholas R J, Emeny M T, Buckle L, Gilbertson A M, Buckle P D and Ashley T 2009 *Phys. Rev.* B **80** 125328

[20] Kallaher R L, Heremans J J, Goel N, Chung S J and Santos M B 2010 *Phys. Rev.* B **81** 075303

[21] Vakili K, Shkolnikov Y P, Tutuc E, De Poortere E P and Shayegan M 2004 *Phys. Rev. Lett.* **92** 226401

[22] Shashikin A A, Rahimi M, Anissimova S, Kravchenko S V, Dolgopolov V T and Klapwijk T M 2003 *Phys. Rev. Lett.* **91** 046403

[23] Perez F, Aku-leh C, Richards D, Jusserand B, Smith L C, Wolverson D and Karczewski G 2007 *Phys. Rev. Lett.* **99** 026403

[24] Dingle R B 1952 *Proc. R. Soc. A* **211** 517

[25] Liu H W, Yang K F, Mishima T D, Santos M B and Hirayama Y 2010 *Phys. Rev.* B **82** 241304 (R)

[26] Tutuc E, Melinte S, Poortere E P, Shayegan M and Winkler R 2003 *Phys. Rev.* B **67** 241309 (R)

[27] Das Sarma S and Hwang E H 2000 *Phys. Rev. Lett.* **84** 5596

[28] Leadley D R, Nicholas R J, Harris J J, Foxon C T, 1998 *Phys. Rev.* B **58** 13036

[29] Gilbertson A M, Branford W R, Fearn M, Buckle L, Buckle P D, Ashley T and Cohen L F 2009 *Phys. Rev.* B **79** 235333

[30] Piot B A, Maude D K, Henini M, Wasilewski Z R, Friedland K J, Hey R, Ploog K H, Toropov A I, Airey R and Hill G, 2005 *Phys. Rev.* B **72** 245325

[31] Ando T 1974 *J. Phys. Soc. Jpn* **37** 1044

**Figure 1.** (a) Longitudinal resistance $R_{xx}$ vs. the perpendicular component $B_{perp}$ of the total magnetic field $B$ at varied tilt angles with temperature $T$ = 200 mK and source-drain current of 3 nA. The tilt angle $\theta$ is defined in the inset. (b) Peak positions of $R_{xx}$ (solid dots) vs. $B$ and $B_{perp}$ (or filling factor $v$). The dash-dot line guides the Landau-level (LL) intersection at integer coincidence ratio $r$. The hollow symbols denote $R_{xx}$ minima and maxima at coincidence angle $\theta_c$ and integer filling factor $v$. The thick solid line accentuates the resistance spike at the LL intersection. In an angle range between 85.3º and 89.7º (so-called parallel-field region), the solid square marks the critical field $B_c$ for full spin polarization and the thin solid line guides to the eyes. The inset schematically shows the LL evolution in tilted fields for constant cyclotron energy $E_c$. Integers inside the diagram denote $v$.

**Figure 2.** (a) $R_{xx}$ vs. $B$ at $v$ = 6 and $v$ = 7 obtained from the raw data of figure 1(b). (b) $R_{xx}$ vs. $B$ in the parallel-field region. The dashed line marks the critical field $B_c$ for full spin polarization. A diagram of energy levels in parallel fields is shown in the inset, where $E_F$ denotes the Fermi level. Split of each Zeeman sublevel originates from Landau quantization.

**Figure 3.** (a) Spin polarization $P$ vs. the normalized field $B_p/B_c$ calculated from figure 1(b). (b) $\chi_s$ and $\chi_{gm}$ vs. $P$. The open and solid symbols represent the data obtained from the InSb 2DESs with doped barriers on both sides and one side of the well, respectively. The thick (dashed) line in (a) is the fit to the open (solid) symbols using (3). The thick (dashed) line in the $P$-$\chi_{gm}$ plots of (b) is the fit to the open (solid) symbols using (2). The $P$-$\chi_s$ plots of the samples with two-side (solid line) and one-side (dashed line) doped barriers are calculated from (5).

**Figure 4.** (a) Dingle plots of the temperature ($T$)-dependent amplitude ($A$) of the first two clearly resolved SdH oscillations at $\theta$ = 0°, from which we attain the mass values $m_1$ and $m_2$. (b) Dependence of the effective mass $m^*$ = $(m_1 + m_2)/2$ on $\theta$. The dashed line is a guide for the eye.

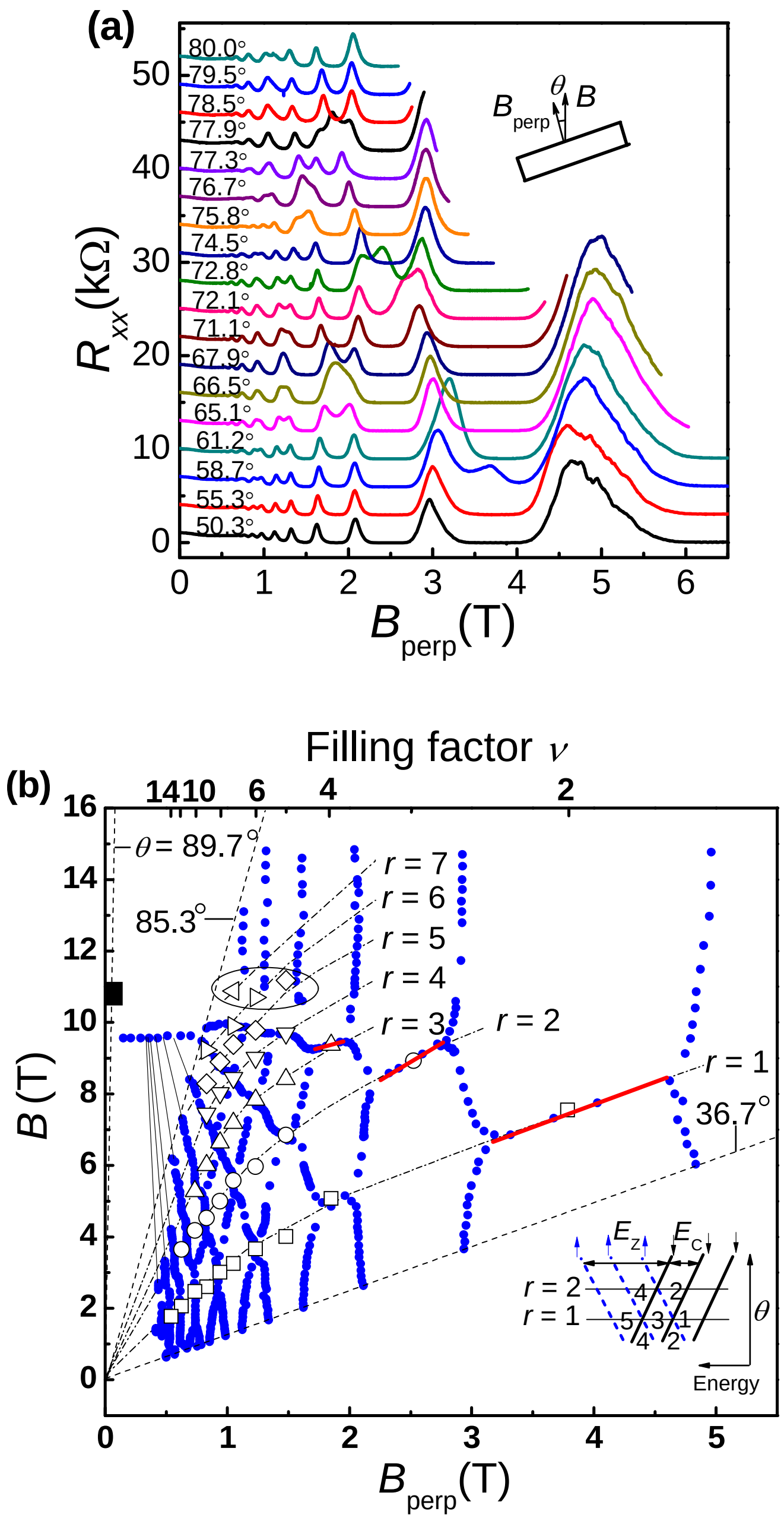

Figure 1

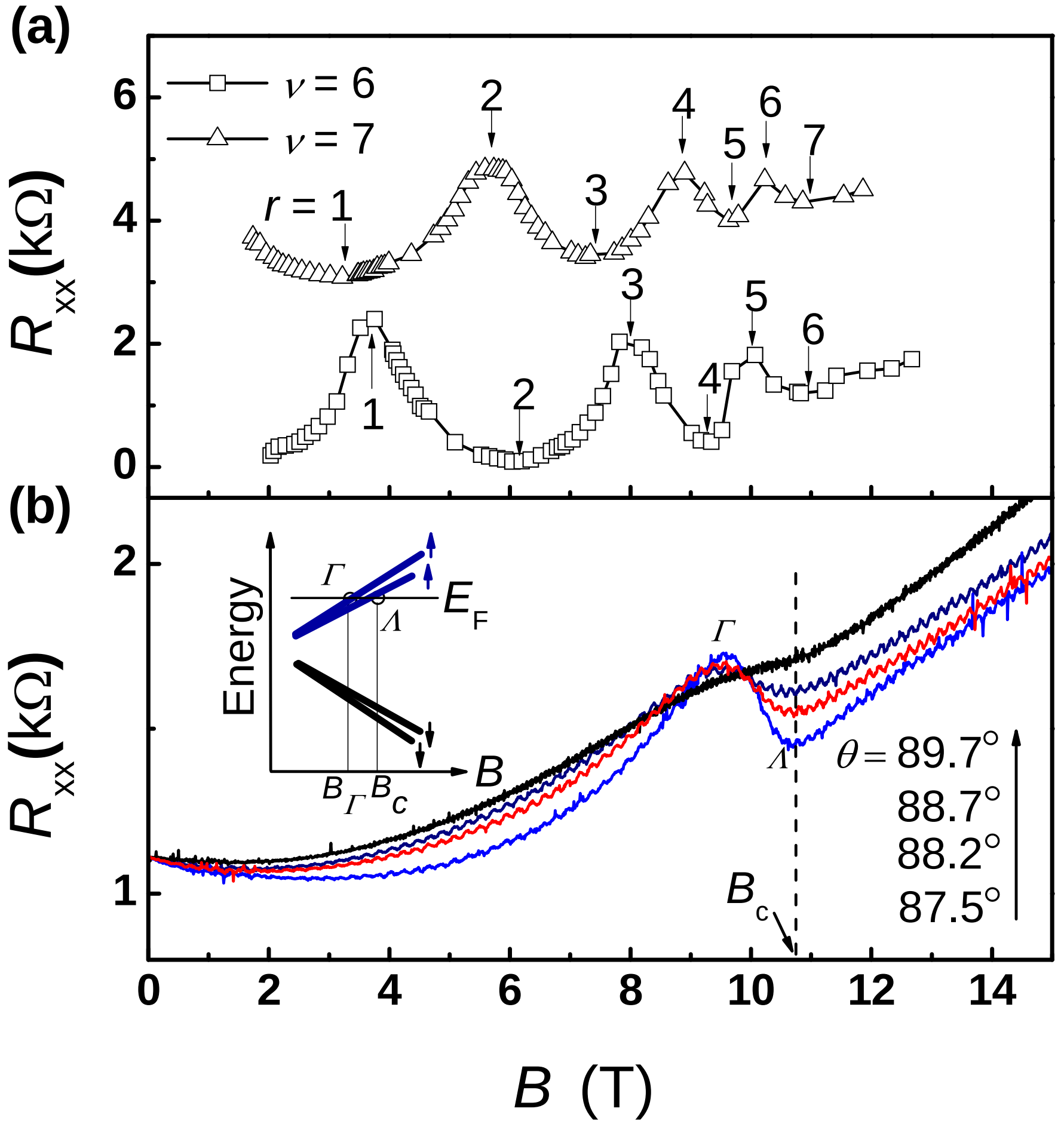

Figure 2

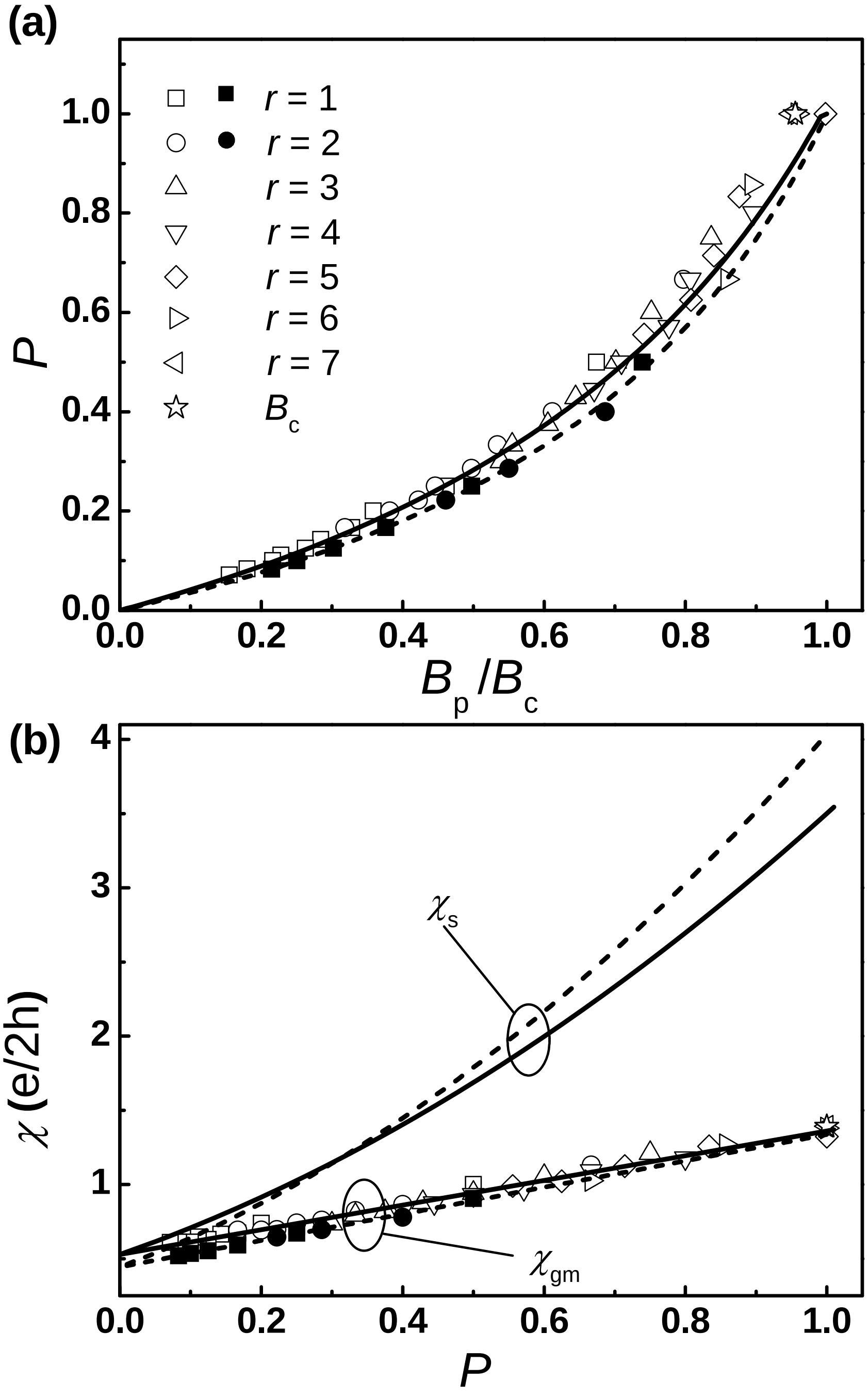

Figure 3

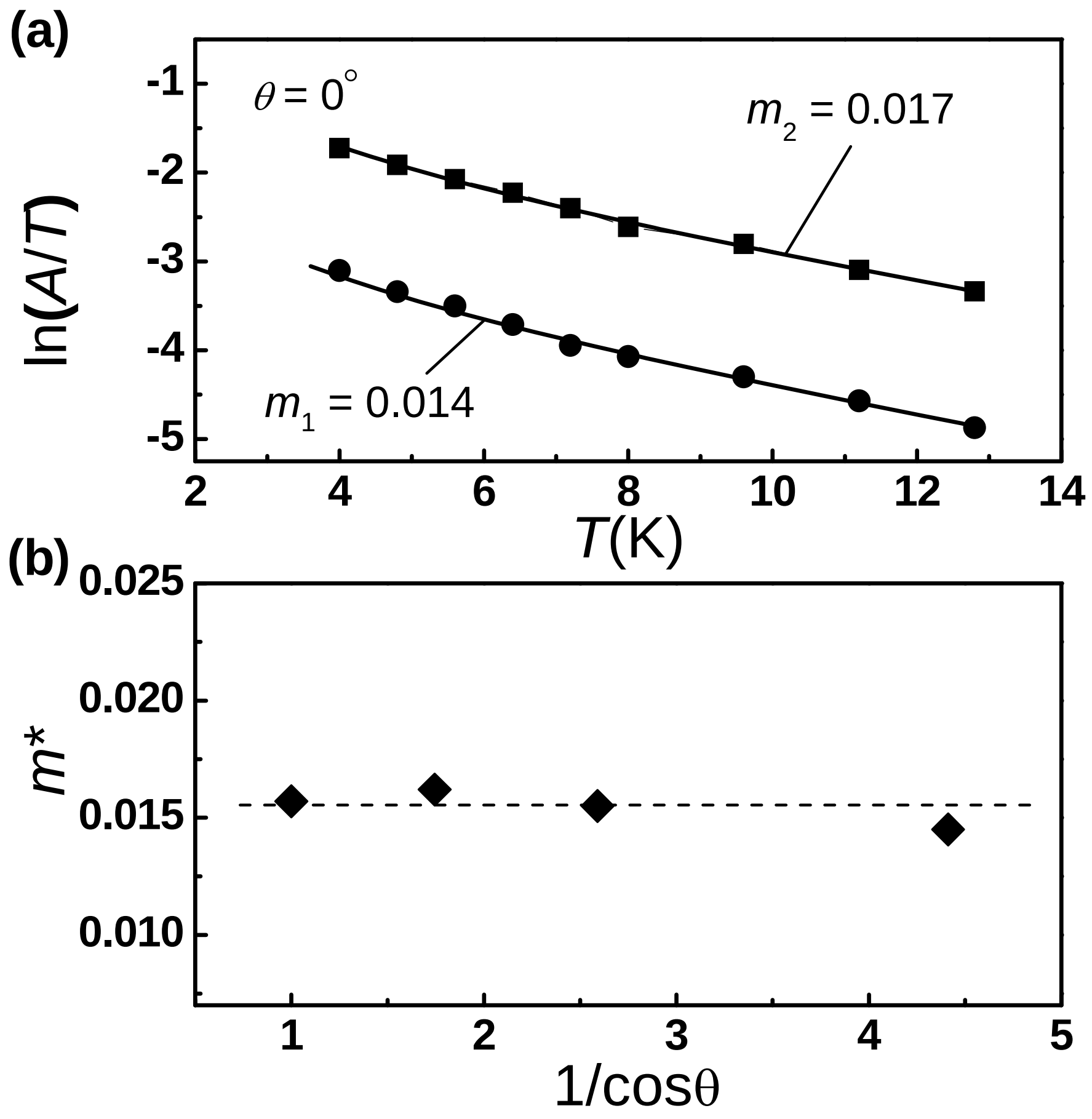

Figure 4